\documentclass[letterpaper, twocolumn,iop]{emulateapj}
\shorttitle{AO Imaging of KOIs}
\shortauthors{Dressing et al. 2014}
\usepackage{epsfig}
\usepackage{natbib}
\usepackage{array}
\usepackage{enumerate}
\usepackage{amsmath}
\usepackage{rotating}
\usepackage{url}
\usepackage{longtable}
\newcommand{\cN}[1]{\mathcal{N}}

\def\gsim{\;\rlap{\lower 2.5pt
 \hbox{$\sim$}}\raise 1.5pt\hbox{$>$}\;}
\def\lsim{\;\rlap{\lower 2.5pt
   \hbox{$\sim$}}\raise 1.5pt\hbox{$<$}\;}

\def\rearth{{\rm\,R_\oplus}}		   
\def\kepler {{\emph{Kepler}\,}}

\newcommand{\ksmag}{\emph{Ks}~magnitudes }
\newcommand{\jmag}{\emph{J}~magnitudes }

\newcommand{\kpmag}{\emph{Kp}~magnitude }
\def\nao{{87}} 
\def\nfp{9} 
\def\nhosts{81} 
\def\nsampcp{34} 
\def\nsamppc{92} 
\def\nsampfp{9} 

\def\nkois{4234} 
\def\nconfplan{977} 
\def\ncomp{27} 
\def\nwithinone{five} 
\def\nwithintwo{seven} 
\def\nwithinfour{15} 

\def\percentwithintwo{13} 

\def\tabstardist{10} 

\begin{document}


\title{Adaptive Optics Images III: 87 \emph{Kepler} Objects of Interest\altaffilmark{*}}
\author{Courtney D. Dressing\altaffilmark{1,2}}
\author{Elisabeth R. Adams\altaffilmark{3}}
\author{Andrea K. Dupree\altaffilmark{1}}
\author{Craig Kulesa\altaffilmark{4}}
\author{Don McCarthy\altaffilmark{4}}
\altaffiltext{*}{Observations reported here were obtained at the MMT Observatory, a joint facility of the Smithsonian Institution and the University of Arizona.}
\altaffiltext{1}{Harvard-Smithsonian Center for Astrophysics, 60 Garden St., Cambridge, MA 02138}
\altaffiltext{2}{{\tt cdressing@cfa.harvard.edu}}
\altaffiltext{2}{Planetary Science Institute, 1700 East Fort Lowell,
Suite 106, Tucson, AZ 85719}
\altaffiltext{4}{Steward Observatory, The University of Arizona, 933 N. Cherry Ave., Tucson, AZ 85721, USA}
 
\date{\today} 


\begin{abstract}
The \kepler mission has revolutionized our understanding of exoplanets, but some of the planet candidates identified by \kepler may actually be astrophysical false positives or planets whose transit depths are diluted by the presence of another star. Adaptive optics images made with ARIES at the MMT of \nao~\kepler Objects of Interest place limits on the presence of fainter stars in or near the \kepler aperture. We detected visual companions within $1''$ for \nwithinone~stars, between $1''$ and $2''$ for \nwithintwo~stars, and between $2''$ and $4''$ for \nwithinfour~stars. For those systems, we estimate the brightness of companion stars in the \kepler bandpass and provide approximate corrections to the radii of associated planet candidates due to the extra light in the aperture. For all stars observed, we report detection limits on the presence of nearby stars. ARIES is typically sensitive to stars approximately $5.3~\emph{Ks}$ magnitudes fainter than the target star within $1''$ and approximately $5.7~\emph{Ks}$ magnitudes fainter within $2''$, but can detect stars as faint as $\Delta Ks = 7.5$ under ideal conditions.
\end{abstract}
\maketitle

\section{Introduction}

Since launch in 2009, the \kepler mission has discovered \nkois~planet candidates and confirmed or validated \nconfplan~planets \citep{borucki_et_al2010, borucki_et_al2011a, borucki_et_al2011b, batalha_et_al2013, burke_et_al2014, rowe_et_al2014}. Many of the planet candidates are expected to be bona fide planets \citep{borucki_et_al2011a, morton+johnson2011, fressin_et_al2013}, but a small fraction may actually be astrophysical false positives \citep{brown2003} in which the apparent transit signal is produced by a pair of eclipsing stars physically associated with the target star (a hierarchical triple) or in the background of the target star (a background eclipsing binary). Close-in giant planet candidates \citep{santerne_et_al2012} and candidate planets around giant stars \citep{sliski+kipping2014} are particularly likely to be false positives. In other cases, transit signals may be diluted due to the presence of other stars (physically associated or not) in the target aperture. This would cause the radius of the planet to be underestimated. Because \kepler has a relatively large plate scale of nearly 4$''$ per pixel and many target apertures consist of multiple pixels, acquiring higher resolution follow-up imagery near planet host stars is crucial for untangling potentially blended systems. 

In order to reduce the odds of classifying blended systems as planet candidates, the \kepler team performs a series of tests on \kepler Objects of Interest (KOIs) before nominating them to planet candidate status. The tests include comparing the depths of odd and even transits in order to identify stellar eclipses that have been misidentified as planetary transits, checking for ellipsoidal variations, and searching for secondary eclipses \citep{batalha_et_al2010}. Background eclipsing binaries can also be identified by examining the direction and magnitude of the shift in the photocenter during transit \citep{bryson_et_al2013}. Even systems in which the observed dip is due to a transit on the target star can have significant centroid motion in crowded fields. However, some background eclipsing binaries can be identified by computing the ``source offset'' between the target star and the transit source \citep{bryson_et_al2013}. A dip due to a transit on the target star should have a negligible source offset while a dip due to a transit or eclipse of another star can  result in a significant source offset depending on the angular separation and relative brightnesses of the true source and the target star.

An additional false positive check that can be performed using \kepler data alone consists of a comparison of the ephemeris of an identified transit signal to the ephemerides of known eclipsing binaries, variable stars, and other planet candidates. Using this method and supplementing \kepler data with additional catalogs of eclipsing binaries and variable stars, \citet{coughlin_et_al2014} found that 12\% of the KOIs they inspected were false positives due to contamination from other known sources.  However, because \kepler does not downlink data for all stars in the field, some contaminated KOIs will not be revealed via ephemeris matching because the contaminating star will not be downloaded. Correcting for this effect, \citet{coughlin_et_al2014} caution that 35\% of KOIs may be false positives due to contamination. 

In an ideal case, all planet candidates could be confirmed by obtaining radial velocity observations and measuring a mass for the transiting planet. However, this plan is logistically impossible due to the large number of planet candidates, the faint magnitudes of most \kepler host stars, and the small RV signature expected for most small planets. In many cases, we must therefore attempt to ``validate'' planet candidates by demonstrating that the odds that the transit signal is due to a bona fide transiting planet are much higher than the odds of a false positive \citep[e.g.,][]{ballard_et_al2011, cochran_et_al2011, torres_et_al2011, fressin_et_al2011, borucki_et_al2012, ford_et_al2012, fressin_et_al2012, lissauer_et_al2012, morton2012, ballard_et_al2013, lissauer_et_al2014, rowe_et_al2014, wang_et_al2014}. 

In order to validate planets and properly correct for diluted transits, we need to place limits on the presence of other stars close to the target. The Kepler-14 system is a prime example of the importance of high-resolution imaging: \citet{buchhave_et_al2011} report that the planetary radius and mass would have been underestimated by 10\% and 60\%, respectively, without high-resolution follow-up images. Their analysis of ground-based follow-up images revealed that the target star is in a close binary system with a nearby star only 0.5~magnitudes fainter and 0\farcs3 away. The \kepler team and community have used speckle imaging \citep{howell_et_al2011, horch_et_al2012, kane_et_al2014}, lucky imaging \citep{lillo-box_et_al2012}, high-resolution adaptive optics imaging \citep{adams_et_al2012, adams_et_al2013a, adams_et_al2013b, law_et_al2013}, and Hubble Space Telescope snapshots (SNAP Program 12893; PI: R. Gilliland) to accomplish this objective. 

In this paper, we present adaptive optics images of \nao~\kepler planet candidates acquired in September~2012, October~2012, and September~2013 in order to investigate whether any of the target stars are diluted due to nearby stars and to place limits on the presence of additional stars in the \kepler target aperture. We explain our observing strategy in Section~\ref{sec:obs}, the target sample in Section~\ref{sec:sample}, and our data reduction process in Section~\ref{sec:data}. We then discuss the detected visual companions in Section~\ref{sec:comp} and place limits on undetected stars in Section~\ref{sec:lim}. We compare our findings to the results of previous surveys in Section~\ref{sec:disc} and conclude in Section~\ref{sec:conc}.

\section{Observations}
\label{sec:obs}
All observations were taken using the Arizona Infrared Imager and Echelle Spectrograph (ARIES) behind the adaptive optics system on the 6.5m Multiple-Mirror Telescope (MMT). We used the target star as a natural guide star and ran the AO system at speeds between 10 and 550~Hz depending on the brightness of the target star and the current observing conditions. The resulting full-width at half-maximum of the target star point spread functions (PSFs) varied between 0\farcs1 and 0\farcs58, with a median value of 0\farcs25. The airmass of our targets ranged from 1.01 to 2.01, with a median value of 1.16.

We observed all targets using a four-point dither pattern in $f$/30 mode with a plate scale of $0''.02085$ pixel$^{-1}$ and a field of view of $20'' \times 20''$. We also observed KOI~886 in $f$/15 mode with a plate scale of $0''.04$ pixel$^{-1}$ and a field of view of $40'' \times 40''$, but we opted to use the images taken in $f$/30 mode for the final reduction. The field rotator was turned on for the observations acquired in 2012 but not for the observations taken in 2013. Accordingly, more distant stars are smeared by field rotation in the images acquired in 2013.

Under ideal conditions, ARIES is diffraction-limited in $J,H,Ks$ in $f/30$ mode down to a limiting magnitude of $Ks = 21$, and diffraction-limited in $Ks$ in $f/15$ mode down to a limiting magnitude of $Ks = 22$. The measured Strehl ratios were 0.3 in $Ks$ and 0.05 in $J$ for ARIES observations acquired in May 2010 under favorable observing conditions with uncorrected seeing of $0\farcs5$ in Ks \citep{adams_et_al2012}.

We varied the integration times for individual exposures between 0.8~seconds and 89.9~seconds depending on the stellar magnitude. Our observing strategy was intentionally more sensitive to fainter companions around fainter target stars because the amount of transit depth dilution is governed by the brightness ratio of the target star and the contaminating star. For our shortest exposure times of 0.8~seconds, we were sensitive to companions as faint as $Ks = 15.4 - 17.2$ depending on the observing conditions. 

We typically repeated the four-point dither pattern four times for a total of 16~images in $Ks$ band per star. For most targets, the dither pattern had a throw of $2''$, but we increased the throw to $3''$ when we noticed nearby stars in the acquisition image. In nine cases, we also imaged objects with close companions in $J$ band in order to determine the color of the companion and better estimate the relative contribution of each star to the flux measured by \emph{Kepler}. Table~\ref{tab:companions} provides a list of target stars with detected companions. 

\section{Target Sample}
\label{sec:sample}
We conducted our observations as part of the \kepler team follow-up effort. We selected our targets from the lists of \kepler planet candidates available at the time of our observing runs in Fall 2012 and Fall 2013. Some of the planet candidates associated with our target stars were later reclassified as false positives and additional planet candidates were detected in several systems. When selecting our sample, we prioritized relatively bright ($Kp \lesssim 14$) stars with small planet candidates. The $Ks$ magnitude of our selected sample extends from $Ks = 8.6$ to $Ks = 12.7$ with a median magnitude of $Ks = 11.6$.

\section{Data Analysis}
\label{sec:data}
We reduced the ARIES observations of each star using the IRAF and python pipeline described in \citet{adams_et_al2012, adams_et_al2013a, adams_et_al2013b}. We calibrated each set of dithered images using standard IRAF procedures\footnote{\url{http://iraf.noao.edu}} and then used the \emph{xmosaic} function in the \emph{xdimsum} package to combine and sky-subtract the images. For targets with detected companions, we determined the approximate orientation of the field from the dither pattern. Our field orientations are therefore approximate and should be treated as general guidelines with an accuracy of a few degrees.

We searched for visual companions to our target stars by looking for bright objects in the reduced images using the IRAF routine \emph{daophot} and by visually inspecting each image. The automated IRAF routines sometimes triggered on residual PSF speckles and image artifacts near the edges of the CCD and in a square pattern one-half CCD width away from bright stars, but those artifacts were easier to identify visually. Due to the quasi-static nature of speckles, we saw similarities in the speckle pattern throughout the course of the night. We could therefore distinguish between speckles and visual companions by whether the objects reappeared in images of multiple target stars or whether they were unique to a particular target star. 

For the companions that passed visual inspection, we measured the magnitude difference relative to the target star using the IRAF routine \emph{phot}. We adopted a 5~pixel aperture in order to sample most of the PSF of the target star without contaminating the measurement with light from nearby stars. We tested the effect of using larger apertures for stars observed in poor seeing conditions and found only slight changes (0.001 - 0.03~magnitudes) in the differential photometry.

For the closest companions (stars within 0\farcs5 of the target star), we instead determined the relative magnitude by simultaneously fitting the PSFs using the same Mathematica routines as in \citet{adams_et_al2012, adams_et_al2013a, adams_et_al2013b}. Our PSF fitting routine fits a Bessel-Lorentzian-Fourier model to each star using six Bessel and four Fourier terms.

We followed the procedure outlined in  \citet{adams_et_al2012, adams_et_al2013a, adams_et_al2013b} to determine the approximate \kepler magnitude, $Kp$, of identified companions. We first measured the brightness differential between the target stars and companions in $Ks$ (and $J$ when available) and converted those to apparent magnitudes for the companions using the target star $Ks$ and $J$ magnitudes reported in the Two Micron All Sky Survey (2MASS) catalog \citep{skrutskie_et_al2006} as absolute references. For systems with detected companions within $2''$ we assumed that the stars would have been blended in 2MASS. In those cases, we recomputed the magnitudes of each component so that the total system magnitudes were equal to the catalog values. We then estimated the $Kp$ magnitude of companion stars using the relations provided in Appendix A of \citet{howell_et_al2012}. 

\section{Visual Companions}
\label{sec:comp}
\begin{figure*}[htbp]
\begin{center}
\centering
\includegraphics[width=1\textwidth]{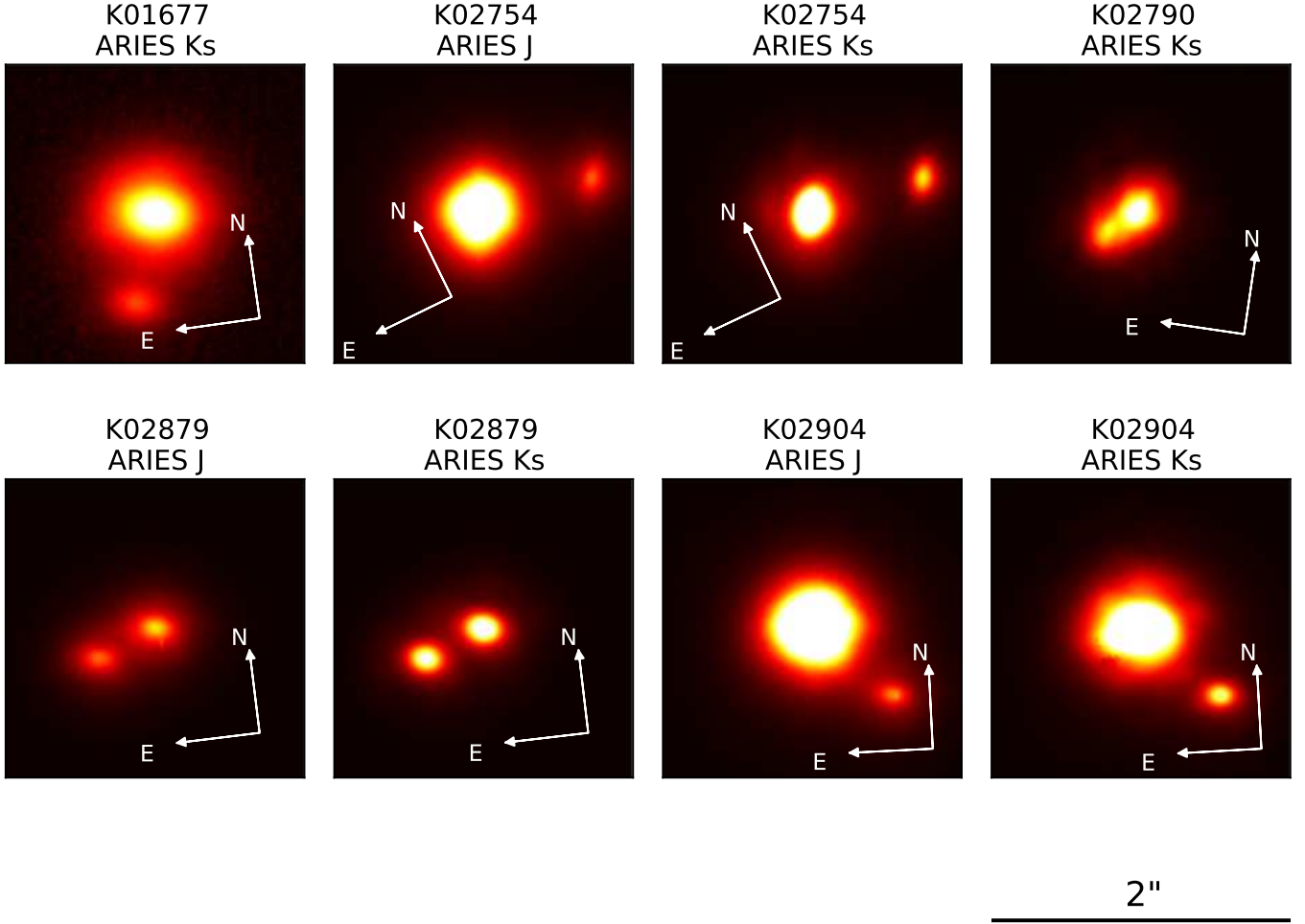}
\end{center}
\caption{Target stars with detected companions within $1''$.  Each box is $2''$ by $2''$. The color scaling is logarithmic for K01677 and linear for the other stars. Only a subset of stars were imaged in both $J$ and $Ks$. Here and in Figures \ref{fig:compin2} and  \ref{fig:compin4} we include all available $J$-band images. }
\label{fig:compin1}
\end{figure*}

\begin{figure*}[htbp]
\begin{center}
\centering
\includegraphics[width=1\textwidth]{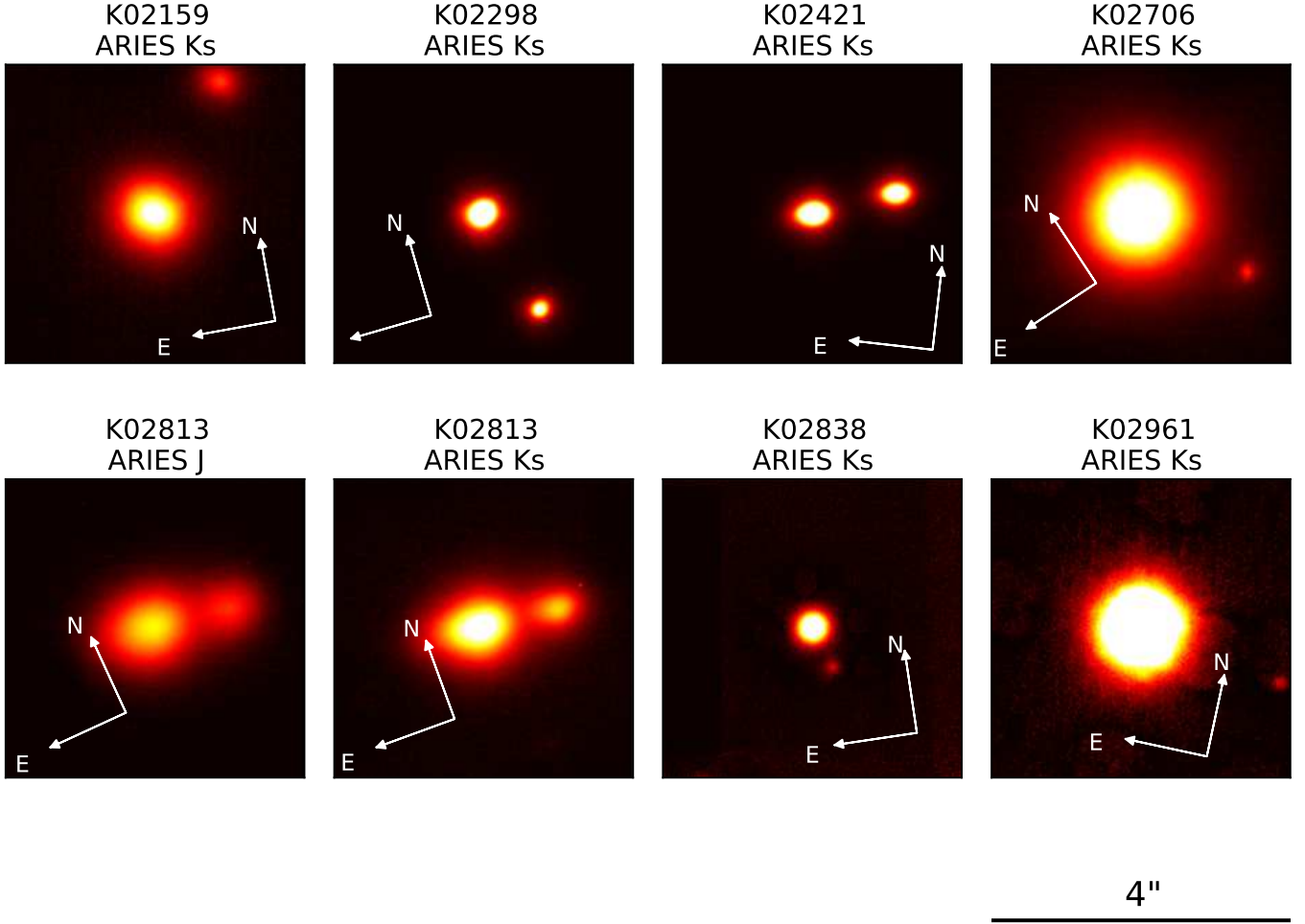}
\end{center}
\caption{Target stars with detected companions between $1''$ and $2''$. Each box is $4''$ by $4''$. The color scaling is linear for K02298 and K02421 and logarithmic for all other stars.}
\label{fig:compin2}
\end{figure*}

\begin{figure*}[htbp]
\begin{center}
\centering
\includegraphics[width=0.9\textwidth]{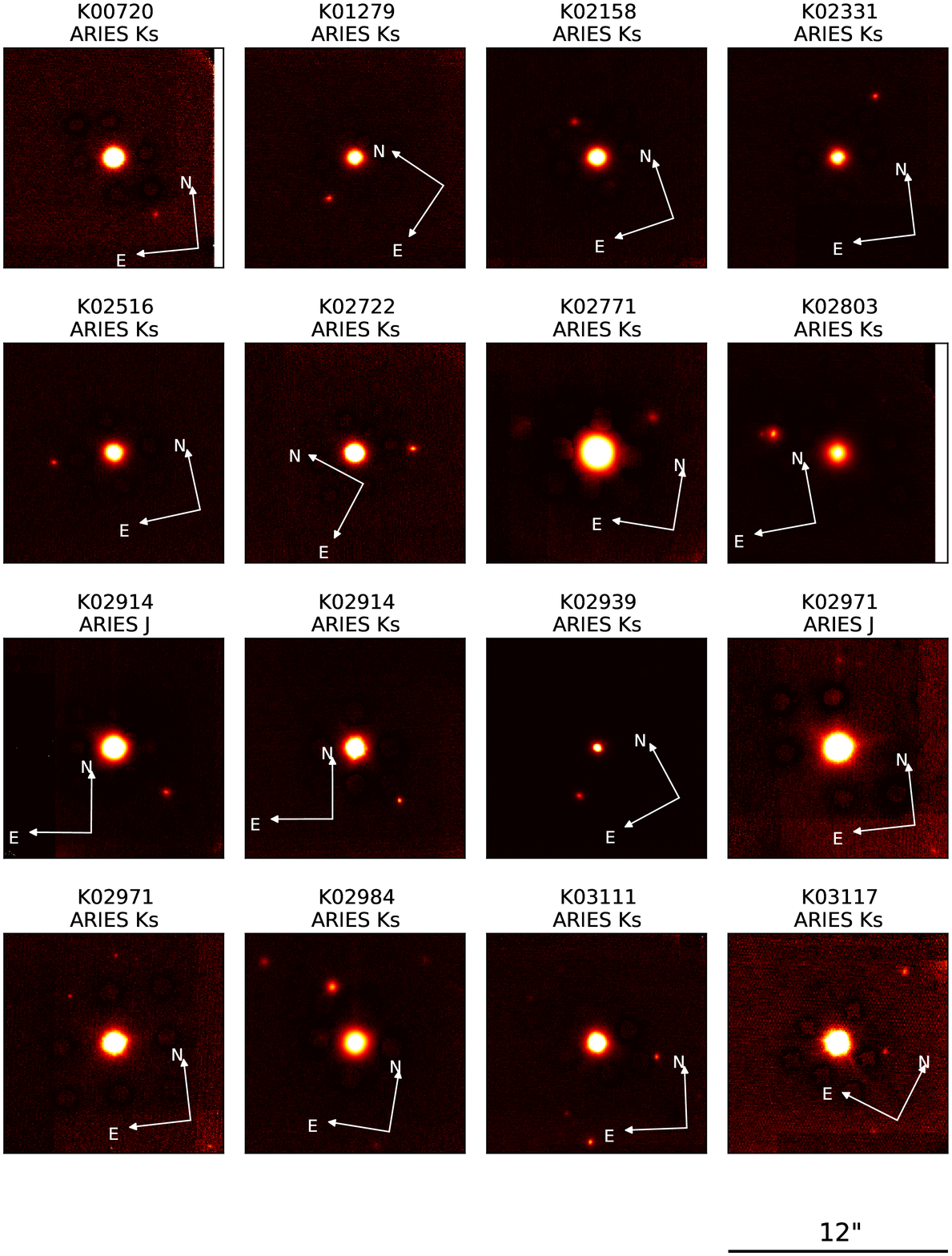}
\end{center}
\caption{Target stars with detected companions between $2''$ and $4''$. Each box is $12''$ by $12''$. KOI~266 also has a companion within $4''$, but it is not pictured here because the companion is smeared due to field rotation. The color scaling is linear for K02939 and logarithmic for all other stars. Some stars also have more distant companions at separations between $4''$ and $12"$. We provide a list of companions within $10''$ in Table~\ref{tab:companions}.}
\label{fig:compin4}
\end{figure*}

Out of \nao~targets, we detected close visual companions for \ncomp~stars: \nwithinone~stars have detected companions within $1''$, \nwithintwo~have detected companions within $2''$, and \nwithinfour~have detected companions within $4''$. We present ARIES images of the stars with companions within $1''$ in Figure~\ref{fig:compin1}, companions within $1-2''$ in Figure~\ref{fig:compin2}, and companions within $2-4''$ in Figure~\ref{fig:compin4}. The ARIES field of view extends to $20''\times20''$, but objects within $4''$, the size of a \kepler pixel, are most likely to dilute planetary transits without revealing their presence by inducing a significant centroid shift. 

For stars with detected companions, the properties of the associated planet candidates will need to be reevaluated to account for the contaminating light in the aperture. We provide rough dilution corrections to the reported planet radii for stars with companions at separations $<2''$. This dilution correction will increase the radii of associated planets by a given percentage. Stars at larger separations also contribute to the background flux due to the large size of \kepler's target apertures, but the fraction of companion star flux collected depends on the \kepler pixel response function, which varies across the focal plane \citep{bryson_et_al2010}, and the specific aperture selected for the target each quarter as the spacecraft is rotated. 

A thorough analysis of the quarter-by-quarter dilution correction for each KOI is beyond the scope of this paper, so we restrict our dilution corrections to a simple order-of-magnitude estimate for the closest companions. In our simple model, we assume that the transit source orbits the target star and that all of the light from both the target and the close companion is captured in the target aperture. In some cases, the transit source might actually be the fainter star detected via adaptive optics imaging rather than the target star. If the planet candidate actually orbits the fainter star, then the planet properties must be completely reevaluated based on the properties of the fainter star. This can result in significant changes to the assumed planet radius, particularly if the fainter star is a background star and not physically associated with the target star. We discuss individual target stars with identified companions in the following sections and list all detected stars within $\tabstardist''$ in Table~\ref{tab:companions}. We caution that this list may be incomplete at larger angular separations because some stars may have been off the edge of the ARIES detector.

\LongTables
 \begin{deluxetable*}{ccrrcccccccrcc}
\tablecolumns{14}
\tabletypesize{\footnotesize}
\tablecaption{Observed Stars with Visual Companions within  $\tabstardist''$}
\tablehead{
\multicolumn{4}{c}{Target Star} &
\multicolumn{4}{c}{} &
\multicolumn{6}{c}{Additional Stars} \\
\noalign{\smallskip}
\cline{0-3}
\cline{9-14}
\noalign{\smallskip}
\colhead{KOI} &
\colhead{\kepler ID} &
\colhead{\emph{Kp\tablenotemark{a}}} &
\colhead{2MASS} &
\colhead{CP\tablenotemark{b}} &
\colhead{PC\tablenotemark{c}} &
\colhead{ND\tablenotemark{d}} &
\colhead{FP\tablenotemark{e}} &
\colhead{Star} &
\colhead{Dist} &
\colhead{Dist Err} &
\colhead{P.A.\tablenotemark{f}} &
\colhead{$\Delta$mag\tablenotemark{g}} &
\colhead{\emph{Kp}} \\
\colhead{} &
\colhead{} &
\colhead{} &
\colhead{\emph{Ks}} &
\colhead{} &
\colhead{} &
\colhead{} &
\colhead{} &
\colhead{} &
\colhead{($''$)} &
\colhead{($''$)} &
\colhead{(\degr)} &
\colhead{(\emph{Ks})} &
\colhead{}
}
K00266	& 7375348	& 11.472	& 10.379	& 0	& 2	& 0	& 0	&1	&3.621	&0.0036	&325.6	&6.32	&19.5\tablenotemark{s} \\
K00364	& 7296438	& 10.087	& 8.645	& 0	& 1	& 0	& 0	&1	&6.019	&0.0015	&130.6	&7.43	&18.7\tablenotemark{s} \\
K00720	& 9963524	& 13.749	& 11.900	& 4	& 0	& 0	& 0	&1	&3.861	&0.0042	&210.7	&5.13	&19.9 \\
...	&...           	&...	&...	&...	&...	&...	&...	&2	&9.047	&0.0019	&76.3	&3.91	&18.3 \\
K01279	& 8628758	& 13.749	& 12.246	& 0	& 2	& 0	& 0	&1	&2.625	&0.002	&91.7	&3.74	&18.6 \\
K01344	& 4136466	& 13.446	& 12.001	& 0	& 1	& 0	& 0	&1	&4.022	&0.0038	&145.0	&4.85	&19.7 \\
K01677	& 5526717	& 14.279	& 12.687	& 0	& 2	& 0	& 0	&1	&0.592	&0.001	&160.5	& 2.48	& 17.6\tablenotemark{P} \\
K01977	& 9412760	& 14.028	& 11.551	& 2	& 0	& 0	& 0	&1	&9.786	&0.0016	&306.1	&5.12	&19.5 \\
K02158	& 5211199	& 13.052	& 11.279	& 0	& 2	& 0	& 0	&1	&2.268	&0.0024	&13.5	&4.45	&18.2 \\
...	&...	&...	&...	&...	&...	&...	&...	&2	&6.484	&0.0018	&334.3	&4.24	&17.9 \\
K02159	& 8804455	& 13.482	& 11.982	& 0	& 1	& 0	& 1	&1	&1.952	&0.0012	&323.4	&2.52	&16.7 \\
...	&...	&...	&...	&...	&...	&...	&...	&2	&6.872	&0.0008	&323.6	&1.1	&15.0 \\
K02298	& 9334893	& 13.831	& 11.735	& 0	& 1	& 0	& 1	&1	&1.469	&0.0004	&195.0	&1.3	&15.2 \\
K02331	& 12401863	& 13.467	& 12.065	& 0	& 1	& 0	& 0	&1	&3.884	&0.0012	&321.3	&3.79	&18.4 \\
K02399	& 11461433	& 14.100	& 12.187	& 0	& 1	& 0	& 0	&1	&4.147	&0.002	&355.1	&4.92	&20.0 \\
K02421	& 8838950	& 14.363	& 12.264	& 0	& 1	& 0	& 0	&1	&1.118	&0.0006	&290.3	&0.42	&15.1 \\
...	&...	&...	&...	&...	&...	&...	&...	&2	&4.002	&0.0013	&130.9	&2.62	&17.8 \\
...	&...	&...	&...	&...	&...	&...	&...	&3	&7.772	&0.0022	&45.6	&4.8	&20.7 \\
K02426	& 8081899	& 13.889	& 12.199	& 0	& 1	& 0	& 0	&1	&8.757	&0.0012	&150.3	&2.58	&16.9 \\
K02516	& 7294743	& 13.388	& 11.582	& 0	& 1	& 0	& 0	&1	&3.306	&0.0014	&86.9	&4.23	&18.3 \\
K02527	& 7879433	& 14.131	& 11.562	& 0	& 1	& 0	& 0	&1	&7.909	&0.0017	&60.1	&3.55	&17.4 \\
K02623	& 10916600	& 13.383	& 12.075	& 0	& 1	& 0	& 0	&1	&5.712	&0.0026	&76.7	&5.39	&20.5 \\
K02672	& 11253827	& 11.921	& 10.285	& 2	& 0	& 0	& 0	&1	&4.541	&0.0018	&294.9	&5.88	&18.8 \\
K02678	& 6779260	& 11.799	& 10.088	& 0	& 1	& 0	& 0	&1	&8.060	&0.002	&132.5	&7.49	&20.7 \\
K02693	& 5185897	& 13.256	& 10.794	& 2	& 1	& 0	& 0	&1	&4.579	&0.0015	&161.0	&5.11	&18.4 \\
K02706	& 9697131	& 10.268	& 9.109	& 0	& 1	& 0	& 0	&1	&1.618	&0.0012	&208.1	&5.2	&16.3 \\
K02722	& 7673192	& 13.268	& 11.993	& 4	& 1	& 0	& 0	&1	&3.151	&0.0025	&32.5	&4.14	&18.7 \\
...	&...	&...	&...	&...	&...	&...	&...	&2	&7.178	&0.002	&200.6	&4.87	&19.7 \\
K02732	& 9886361	& 12.805	& 11.537	& 2	& 2	& 0	& 0	&1	&7.697	&0.0018	&102.8	&6.73	&21.6 \\
...	&...	&...	&...	&...	&...	&...	&...	&2	&9.846	&0.002	&155.7	&6.2	&20.9 \\
K02754	& 10905911	& 12.299	& 10.627	& 0	& 1	& 0	& 0	&1	&0.763	&0.0001	&261.5	&1.65	& 15.3\tablenotemark{P,J} \\
K02771	& 11456382	& 11.751	& 10.462	& 0	& 0	& 0	& 1	&1	&3.574	&0.0012	&312.0	&5.73	&18.8 \\
K02790	& 5652893	& 13.380	& 11.486	& 0	& 1	& 0	& 0	&1	&0.254 	&0.0001	&130.6	 & 0.62	& 14.5\tablenotemark{P} \\
...	&...	&...	&...	&...	&...	&...	&...	&2	&5.662	&0.003	&240.8	&5.39	&20.4 \\
...	&...	&...	&...	&...	&...	&...	&...	&3	&5.303	&0.0016	&6.8	&4.84	&19.7 \\
...	&...	&...	&...	&...	&...	&...	&...	&4	&8.381	&0.0017	&63.4	&5.28	&20.2 \\
K02803	& 9898447	& 12.258	& 10.642	& 0	& 1	& 0	& 0	&1	&3.650	&0.0005	&63.1	&2.64	&15.1 \\
...	&...	&...	&...	&...	&...	&...	&...	&2	&4.245	&0.002	&65.6	&5.13	&18.3 \\
...	&...	&...	&...	&...	&...	&...	&...	&3	&8.606	&0.0016	&205.4	&5.18	&18.3 \\
K02813	& 11197853	& 13.586	& 11.514	& 0	& 1	& 0	& 0	&1	&1.038	&0.0011	&263.6	&1.82	& 14.8\tablenotemark{J} \\
K02829	& 6197215	& 12.824	& 11.444	& 0	& 1	& 0	& 0	&1	&9.922	&0.0016	&120.6	&5.76	&20.2 \\
K02833	& 9109857	& 12.599	& 11.123	& 0	& 1	& 0	& 0	&1	&8.674	&0.0006	&94.9	&4.4	&17.9 \\
...	&...	&...	&...	&...	&...	&...	&...	&2	&7.216	&0.0013	&358.4	&5.42	&19.3 \\
K02838	& 6607357	& 13.421	& 11.857	& 0	& 1	& 0	& 1&1	&1.748	&0.0053	&197.2	&4.04	&18.5 \\
...	&...	&...	&...	&...	&...	&...	&...	&2	&7.739	&0.0018	&137.3	&2.83	&16.8 \\
K02840	& 6467363	& 13.884	& 12.261	& 0	& 2	& 0	& 0	&1	&4.033	&0.0021	&297.0	&4.16	&19.1 \\
...	&...	&...	&...	&...	&...	&...	&...	&2	&7.029	&0.0033	&258.1	&5.05	&20.3 \\
...	&...	&...	&...	&...	&...	&...	&...	&3	&8.221	&0.0018	&185.9	&4.44	&19.5 \\
K02879	& 7051984	& 12.771	& 11.099	& 0	& 0	& 0	& 1	&1	&0.423	&0.0001	&110.9	&0.27	& 13.8\tablenotemark{P,J} \\
...	&...	&...	&...	&...	&...	&...	&...	&2	&5.449	&0.0004	&223.6	& 1.41	&15.0 \\
K02904	& 3969687	& 12.683	& 11.359	& 0	& 1	& 0	& 0	&1	&0.684	&0.001	&226.0	&2.58	& 15.8\tablenotemark{P,J} \\
...	&...	&...	&...	&...	&...	&...	&...	&2	&5.274	&0.0005	&49.5	&2.84	&16.3 \\
...	&...	&...	&...	&...	&...	&...	&...	&3	&8.488	&0.0006	&21.3	&3.36	&17.0\\
K02913	& 9693006	& 12.858	& 11.361	& 0	& 1	& 0	& 0	&1	&7.141	&0.0014	&15.7	&4.42	&18.3 \\
K02914	& 6837283	& 12.199	& 11.006	& 0	& 1	& 0	& 0	&1	&3.740	&0.0012	&220.1	&5.28	&17.8\tablenotemark{J} \\
K02915	& 5613821	& 13.346	& 11.956	& 0	& 1	& 0	& 0	&1	&5.226	&0.0014	&167.3	&5.36	&20.3 \\
...	&...	&...	&...	&...	&...	&...	&...	&2	&9.276	&0.0007	&305.6	&3.82	&18.3 \\
K02939	& 5473556	& 13.545	& 12.006 & 0	&0	& 0	& 0	&1	&2.780	&0.0009	&131.4	& 1.84	&15.8 \\
...	&...	&...	&...	&...	&...	&...	&...	&2	&4.293	&0.0034	&202.7	&5.96	& 21.2 \\
...	&...	&...	&...	&...	&...	&...	&...	&3	&4.820	&0.0036	&300.5	&5.68	& 20.8 \\
...	&...	&...	&...	&...	&...	&...	&...	&4	&9.768	&0.0026	&355.9	&4.16	& 18.8 \\
...	&...	&...	&...	&...	&...	&...	&...	&5	&8.174	&0.0023	&84.3	&4.5	& 19.2 \\
K02961	& 10471515	& 12.581	& 11.290	& 0	& 1	& 0	& 0	&1	&1.954	&0.0021	&260.6	&6.94	&21.5 \\
...	&...	&...	&...	&...	&...	&...	&...	&2	&5.210	&0.0009	&59.9	&4.86	&18.8 \\
K02970	& 5450893	& 12.861	& 11.566	& 0	& 1	& 0	& 0	&1	&4.393	&0.0005	&307.5	&3.16	&16.8 \\
...	&...	&...	&...	&...	&...	&...	&...	&2	&5.802	&0.0024	&279.0	&6.98	&22.0 \\
...	&...	&...	&...	&...	&...	&...	&...	&3	&6.254	&0.0013	&168.9	&6.03	&20.7 \\
K02971	& 4770174	& 12.742	& 11.438	& 0	& 2	& 0	& 0	&1	&3.477	&0.0019	&36.8	&6.69	&21.4 \\
...	&...	&...	&...	&...	&...	&...	&...	&2	&4.736	&0.0017	&352.1	&6.81	&21.6 \\
...	&...	&...	&...	&...	&...	&...	&...	&3	&7.990	&0.0016	&352.5	&6.08	&20.6 \\
...	&...	&...	&...	&...	&...	&...	&...	&4	&6.806	&0.0015	&147.8	&6.98	&21.8 \\
...	&...	&...	&...	&...	&...	&...	&...	&5	&7.680	&0.0014	&216.6	&6.31	&20.9 \\
K02984	& 7918652	& 13.066	& 11.637	& 0	& 1	& 0	& 0	&1	&3.262	&0.001	&31.9	&3.81	&17.8 \\
...	&...	&...	&...	&...	&...	&...	&...	&2	&6.620	&0.0022	&57.8	&5.68	&20.3 \\
...	&...	&...	&...	&...	&...	&...	&...	&3	&5.848	&0.002	&328.6	&6.84	&21.9 \\
...	&...	&...	&...	&...	&...	&...	&...	&4	&5.679	&0.0009	&200.8	&6.46	&21.4 \\
K03015	& 11403530	& 13.219	& 11.774	& 0	& 1	& 0	& 0	&1	&4.763	&0.0014	&252.3	&5.22	&19.9 \\
K03075	& 3328080	& 12.994	& 11.532	& 0	& 1	& 0	& 0	&1	&4.289	&0.0017	&48.8	&7.0	&21.9 \\
K03111	& 8581240	& 12.863	& 11.353	& 0	& 2	& 0	& 0	&1	&3.334	&0.001	&254.5	&5.25	&19.4 \\
...	&...	&...	&...	&...	&...	&...	&...	&2	&5.390	&0.0012	&174.2	&4.86	&18.9 \\
...	&...	&...	&...	&...	&...	&...	&...	&3	&4.178	&0.0015	&154.9	&6.97	&21.7 \\
\pagebreak
K03117	& 6523351	& 13.163	& 11.508	& 0	& 1	& 0	& 0	&1	&2.612	&0.0018	&286.5	&6.1	&20.7 \\
...	&...	&...	&...	&...	&...	&...	&...	&2	&5.294	&0.0035	&343.4	&5.66	&20.1 \\
...	&...	&...	&...	&...	&...	&...	&...	&3	&5.182	&0.0053	&229.0	&6.42	&21.1 \\
K03122	& 12416661	& 12.086	& 10.819	& 0	& 1	& 0	& 0	&1	&4.437	&0.0017	&55.8	&6.53	&20.4 \\
...	&...	&...	&...	&...	&...	&...	&...	&2	&8.771	&0.0015	&114.9	&6.22	&20.0 \\
K03128	& 7609674	& 13.371	& 11.900	& 0	& 1	& 0	& 0	&1	&6.274	&0.0015	&106.3	&5.14	&20.0 \\
K03242	& 6928906	& 12.374	& 11.520	& 0	& 1	& 0	& 0	&1	&3.998	&0.0012	&238.2	&8.4	&23.8\tablenotemark{s}
 \enddata
 \tablenotetext{a}{Apparent magnitude in the \kepler bandpass.}
 \tablenotetext{b}{Number of confirmed planets associated with the target.}
 \tablenotetext{c}{Number of planet candidates associated with the target.}
 \tablenotetext{d}{Number of not dispositioned KOIs associated with the target.}
 \tablenotetext{e}{Number of false positive KOIs associated with the target.}
 \tablenotetext{f}{Angle measured eastward from north. We caution that the position angles were estimated from the dither pattern and therefore might differ from the true angle by a few degrees.}
 \tablenotetext{g}{Error on $\Delta Ks$ is roughly 0.02~mag.}
 \tablenotetext{J}{Estimated $Kp$ for a dwarf companion based on both $J$ and $Ks$ photometry. }
 \tablenotetext{P}{Brightness contrast, separation, and companion $Kp$ determined using PSF fitting.}
\tablenotetext{s}{These companions to K00266, K00364, and K03242 were smeared by field rotation and their magnitudes are likely underestimated. More distant smeared companions to these stars were omitted from this list.}
\label{tab:companions}
\end{deluxetable*}

\subsection{KOI~266}
This system contains a $1.6\rearth$ planet candidate with a 25.3~day period and a second $1.8\rearth$ planet candidate with a 47.7~day period \citep{burke_et_al2014}. Our ARIES observations revealed a star roughly $6.3$~\ksmag fainter than KOI~266 at a distance of 3\farcs62. KOI~266 was previously inspected\footnote{The archival observations discussed in this paper were reported on the \kepler Community Follow-up Observing Program (CFOP) website: \url{http://cfop.ipac.caltech.edu}.} using speckle imagery with DSSI on WIYN, but the nearby star was beyond the $3\farcs2\times3\farcs2$ speckle field of view. This star was previously imaged by \citet{adams_et_al2012, adams_et_al2013b}, who reported that the companion is 6.6~\jmag fainter and 6.1~\ksmag fainter than KOI~266, resulting in an estimated \kepler magnitude of $Kp = 19.3$. For the rest of this paper, we adopt the magnitude estimates from \citet{adams_et_al2012, adams_et_al2013b} because the visual companion is slightly smeared in our image due to field rotation. The nearby star is also listed in UKIRT \citep{lawrence_et_al2007}. 

KOI~266 was classified by \citet{slawson_et_al2011} as a detached eclipsing binary with a period of 25.3~days, suggesting that the $1.6\rearth$ planet candidate with the same period might not actually be a planet. Instead, the observed decrease in flux every 25.3~days might be an eclipsing binary diluted by the light of a nearby star. The centroid source offset during transits of KOI~266.01 is 0\farcs574 ($2.66\sigma$).

\subsection{KOI~720 (\kepler-221)}
This system has four confirmed planets with radii of 2.96, 2.81, 3.05, and $1.56\rearth$ \citep{borucki_et_al2011b, rowe_et_al2014}. We detected another star 3\farcs86 from the target star. The nearby star is 5.13~\ksmag fainter than KOI~720 and is predicted to have $Kp = 19.9$ ($\Delta Kp = 6.2$). KOI~720 has been observed with the Differential Speckle Survey Instrument (DSSI) on WIYN and at low quality with Robo-AO on the Palomar 1.5-m \citep{law_et_al2013}. The companion we detected is visible in UKIRT and has a reported $J$~magnitude of 17.67. The maximum dilution correction for a star 5.13~\ksmag fainter than the target star is a 0.2\% correction to the planet radii, so the dilution correction is unlikely to be significant given the nearly $4''$ separation between KOI~720 and the companion and the large brightness contrast between the stars. 

\subsection{KOI~1279}
This system contains two short-period planet candidates with radii of $1.6\rearth$ and $0.9\rearth$ \citep{borucki_et_al2011b, batalha_et_al2013}. 
We detected a star 3.74~\ksmag fainter than KOI~1279 at a distance of 2\farcs62. The star is predicted to have $Kp = 18.6$ ($\Delta Kp = 4.8$). KOI~1279 has also been observed using speckle imaging with DSSI on WIYN and at low quality with Robo-AO on the Palomar 1.5-m \citep{law_et_al2013}. The companion we detected was visible in the UKIRT image of the field and has a reported $J$ magnitude of $J = 16.54$. KOI~1279 does not exhibit a large source offset during transits, which supports the interpretation that the planet candidates orbit the target star.

\subsection{KOI~1677}
KOI~1677 hosts a $2.2\rearth$ planet candidate with a 52.1~day orbit and a $0.8\rearth$ candidate with a 8.5~day orbit \citep{batalha_et_al2013}. We detected a companion 2.48~\ksmag fainter than KOI~1677 at a distance of 0\farcs6. Using the relation from \cite{howell_et_al2012}, the predicted \kpmag for the companion is $Kp=17.6$ ($\Delta Kp = 3.3$). This object was also detected in a medium-quality Robo-AO image of KOI~1677 and has an estimated magnitude of $i=18.83 \pm 0.44$ \citep{law_et_al2013}. Assuming that all of the flux from the target star and the companion is captured in the \kepler aperture and that the planet orbits the target star, the planet radius estimate should be increased by roughly 2\% to account for the contamination from the nearby star. KOI~1677 does not display a significant offset during the transits of KOI~1677.01, but the centroid analysis for KOI~1677.02 is not yet available. 

\subsection{KOI~2158}
This system contains two planet candidates with periods of 4.6 and 6.7~days and radii of 1.6 and $1.0\rearth$, respectively \citep{batalha_et_al2013}. We detected a companion 4.45~\ksmag fainter than KOI~2158 at a distance of 2\farcs27. The companion is predicted to have $Kp =18.2$ ($\Delta Kp = 5.2$). KOI~2158 has also been observed with DSSI on WIYN and at medium quality with Robo-AO on the Palomar 1.5-m. The companion was detected in UKIRT and has a reported $J$ magnitude of 17.15. KOI~2158 does not exhibit a large source offset during the transits of either planet candidate. 

\subsection{KOI~2159}
\label{ssec:2159}
KOI~2159 hosts one candidate planet with a period of 7.6~days and a radius of $1.1\rearth$ \citep{batalha_et_al2013}. The NASA Exoplanet Archive entry for KOI~2159 also includes a $1\rearth$ false positive at a period of 2.4~days. Our ARIES observations revealed a companion 2.52~\ksmag fainter than KOI~2159 at a distance of 1\farcs95. The estimated $Kp$ magnitude for the companion is $16.7$ ($\Delta Kp = 3.2$). This companion was also listed as a likely detection with Robo-AO in \citet{law_et_al2013} based on a medium-quality image and has an estimated magnitude of $i=17.28 \pm 0.53$.  The star was also detected in UKIRT and has a $J$ band magnitude of 15.57. The estimated dilution correction due to the extra light from the companion is a 3\% increase to the planet radius. KOI~2159 does not display a significant source offset during transit.

\subsection{KOI~2298}
This system contains a $1\rearth$ planet candidate with a 16.7~day orbit \citep{batalha_et_al2013}. NEXSci also reports a $0.8\rearth$ false positive with a 31.8~day period. We detected a companion 1.3~\ksmag fainter than KOI~2298 at a distance of 1\farcs47. The companion is expected to have $Kp = 15.2$ ($\Delta Kp = 1.4$), indicating that the contamination from this companion star may lead to a significant underestimate of the planet radius. In the simple approximation that all light from the companion star is captured in the \kepler aperture, the radius estimated for the planet should be increased by 13\% to account for the dilution if indeed the planet orbits the target star and the companion has $Kp =14.9$. However, the companion may be the same object identified roughly $1''$ away from KOI~2298 in a HIRES guider image\footnote{\url{https://cfop.ipac.caltech.edu/edit_obsnotes.php?id=2298}}. The estimated brightness contrast from the HIRES image is three~magnitudes, which implies the companion is red enough that the dilution correction might be only a 3\% change to the radius of the planet candidate. The false positive KOI~2998.02 failed the centroid test during data validation, but KOI~2298 does not exhibit a significant source offset during the transit of KOI~2298.01.

\subsection{KOI~2331}
KOI~2331 hosts a single $1.4\rearth$ planet candidate with a 2.8~day period \citep{batalha_et_al2013}. Our ARIES observations revealed a companion 3.79~\ksmag fainter than KOI~2331 at a separation of 3\farcs88. The predicted $Kp$ magnitude for the companion is $Kp = 18.4$ ($\Delta Kp = 4.9$). KOI~2331 has also been observed at medium quality with Robo-AO on the Palomar 1.5-m \citet{law_et_al2013}. KOI~2331 does not display a significant source offset during transit. 

\subsection{KOI~2421}
KOI~2421 hosts a $0.7\rearth$ planet candidate with a 2.3~day orbit \citep{batalha_et_al2013}. We detected two companions 0.42 and 2.62~\ksmag fainter than KOI~2421 at separations of 1\farcs12 and 4\farcs0, respectively. The closer companion is predicted to be $Kp = 15.1$ ($\Delta Kp = 0.7$) and the farther companion is predicted to be $Kp = 17.8$ ($\Delta Kp = 3.5$). KOI~2421 has also been observed using NIRC2 on Keck with a laser guide star. 

Due to the similar brightness of the innermost companion and KOI~2421, this system will require a significant dilution correction. If all light from the innermost companion is captured in the \kepler aperture and the planet orbits the target star, then the planet radius measurement will need to be increased by 23\% to account for dilution. KOI~2421 does not exhibit a large source offset during transit, which lends support to the theory that the planet candidate orbits the target star, but this system should be inspected closely to confirm that the planet candidate does indeed orbit KOI~2421.

\subsection{KOI~2516}
KOI~2516 hosts a $1.2\rearth$ planet candidate with a 2.8~day orbit \citep{batalha_et_al2013}. We detected a companion 4.23~\ksmag fainter than KOI~2516 at a distance of 3\farcs31. The estimated \kpmag of the companion star is $Kp = 18.3$ ($\Delta Kp = 4.9$). The companion was previously identified in UKIRT and has a $J$ magnitude of 16.45. KOI~2516 does not display a significant source offset during transit.

\subsection{KOI~2706}
KOI~2706 hosts a $1.5\rearth$ planet candidate with a 3.1~day orbit \citep{burke_et_al2014}.  We detected a companion 5.2~\ksmag fainter than KOI~2706 at a separation of 1\farcs62. The predicted \kpmag for the companion is $Kp = 16.3$ ($\Delta Kp = 6.0$). KOI~2706 has also been observed with DSSI on WIYN and PHARO on the Palomar-5m. The detected companion is visible in the PHARO observations, but was undetected in the WIYN speckle imaging. The estimated dilution correction for this system is a 0.2\% increase in the radius of the planet candidate. KOI~2706 exhibits a 0\farcs83 ($3.7\sigma$) source offset during transit.\footnote{Data Validation reports containing centroid analyses are available at \url{http://exoplanetarchive.ipac.caltech.edu} for all KOIs.}

\subsection{KOI~2722 (\kepler-402)}
This system contains four confirmed planets with radii of 1.4, 1.4, 1.1, and $1.3\rearth$ and one candidate planet with a radius of $1.3\rearth$ \citep{burke_et_al2014}. We detected a companion 4.14~\ksmag fainter than KOI~2722 at a distance of 3\farcs15. The estimated \kpmag of the companion star is $Kp = 18.7$ ($\Delta Kp = 5.5$). 

\subsection{KOI~2754}
KOI~2754 hosts a $0.7\rearth$ planet candidate with a 1.3~day period \citep{burke_et_al2014}. Our ARIES observations revealed a companion 2.11~\jmag and 1.65~\ksmag fainter than KOI~2754 at a separation of 0\farcs763. The predicted \kpmag of the companion star is $Kp=15.3$ ($\Delta Kp=3.0$) if the star is a dwarf and $Kp=15.2$ ($\Delta Kp = 2.9$) if the star is a giant. The companion star was also detected in the WIYN speckle imaging of K02754 acquired with DSSI by Mark Everett\footnote{\url{https://cfop.ipac.caltech.edu/edit_target.php?id=2754}}. The companion is 3.12~magnitudes fainter than KOI~2754 at 692nm and 2.56~magnitudes fainter at 880nm. Assuming that the planet orbits the target star and that all of the light from the companion is captured in the \kepler aperture, then the planet radius should be increased by 3\% to account for dilution. KOI~2754 does not exhibit a significant source offset during transit.

\subsection{KOI~2771}
KOI~2771 was reported to have a $1.7\rearth$ planet with a 0.8~day period, but this signal has been found to be a false positive \citep{burke_et_al2014}. We detected a companion 5.73~\ksmag fainter than KOI~2771 at a separation of 3\farcs57. The estimated \kpmag of the companion is $Kp=18.8$ ($\Delta Kp = 7.1$). KOI~2771 has also been observed with DSSI on WIYN and PHARO on the Palomar-5m. The detected companion is visible in both the PHARO observation and in the UKIRT data. 

\subsection{KOI~2790}
KOI~2790 hosts a $0.9\rearth$ planet candidate with a 14.0~day period \citep{burke_et_al2014}. Our ARIES observations revealed a companion 0.62~\ksmag fainter than KOI~2790 at a separation of 0\farcs21. The predicted \kpmag of the companion is $Kp=14.5$ ($\Delta Kp = 1.1$). The companion is clearly identifiable in the more recent higher resolution image of KOI~2790 acquired with NIRC2 on Keck. The approximate increase to the planet radius is 17\% assuming that all of the light from the companion is captured in the \kepler aperture and that the planet orbits the target star. KOI~2790 does not exhibit a significant source offset during transit.

\subsection{KOI~2803}
KOI~2803 hosts a $0.5\rearth$ planet candidate with a 2.4~day period \citep{burke_et_al2014}. We detected a companion 2.64~\ksmag fainter than KOI~2803 at a distance of 3\farcs65. The estimated \kpmag of the companion is $Kp = 15.1$ ($\Delta Kp = 2.9$). KOI~2803 has also been observed with speckle imaging using DSSI on WIYN, but the companion was too far from the star to be detected. The companion we identified was found in UKIRT at a separation of 3\farcs4 and has a reported $J$ magnitude of 18.33. KOI~2803 does not display a significant source offset during transit.

\subsection{KOI~2813}
KOI~2813 hosts a $1.2\rearth$ planet candidate with a 0.7~day period \citep{burke_et_al2014}. We detected a companion 1.43~\jmag and 1.82~\ksmag fainter than KOI~2813 at a separation of 1\farcs04. The predicted \kpmag of the companion is $Kp = 14.8$ ($\Delta Kp = 1.3$). In the simple approximation in which all of the light from the companion star is captured in the \kepler aperture and the companion orbits the target star, then the estimated planet radius should be increased by 15\% to correct for the extra light in the aperture. KOI~2813 does not exhibit a significant source offset during transit. 

\subsection{KOI~2838}
KOI~2838 hosts a $0.7\rearth$ planet candidate with a 4.8~day period. The \kepler data also revealed a 7.7~day false positive \citep{burke_et_al2014}. We detected a companion 4.04~\ksmag fainter than KOI~2838 at a distance of 1\farcs75. The estimated \kpmag of the companion is $Kp = 18.5$ ($\Delta Kp=5.0$) and the approximate dilution correction is a $0.5\%$ increase to the radius of the planet candidate. KOI~2838 does not display a significant source offset during the transits of KOI~2838.02.

\subsection{KOI~2879}
KOI~2879 was reported to have a $1.4\rearth$ planet with a 0.3~day period \citep{burke_et_al2014}, but this signal has been found to be a false positive. Our ARIES observations revealed a companion 0.37~\jmag and 0.27~\ksmag fainter than KOI~2879 at a distance of 0\farcs423. Using the  $J - Ks$ to $Kp - Ks$ color-color conversion from \citet{howell_et_al2012}, we predict that the \kpmag of the companion is $Kp = 13.8$ ($\Delta Kp = 1.1$) if the star is a dwarf or $Kp = 13.9$ ($\Delta Kp = 1.2$) if the star is a giant. If any additional planet candidates are detected around KOI~2879, their radii will need to be increased by roughly 17\% to account for the additional light in the aperture. 

\subsection{KOI~2904}
KOI~2904 hosts a $1.2\rearth$ planet with a 16.4~day period \citep{burke_et_al2014}. We detected a companion 2.74~\jmag and 2.58~\ksmag fainter than KOI~2904 at a separation of 0\farcs68. Using the  $J - Ks$ to $Kp - Ks$ color-color conversion from \citet{howell_et_al2012}, we predict that the \kpmag of the companion is $Kp = 15.8 $ ($\Delta Kp = 3.1 $) if the star is a dwarf or $Kp = 15.9$ ($\Delta Kp = 3.2 $) if the star is a giant. KOI~2904 has also been observed with speckle imaging using DSSI on WIYN and the companion was detected with magnitude differences of 2.83~mags at 692nm and 2.77~mags at 880nm. The estimated dilution correction for this system is 3\% assuming that the planet orbits the target star and that all of the light from the companion is captured in the \kepler aperture. KOI~2904 does not display a significant source offset during transit.

\subsection{KOI~2914}
KOI~2914 hosts a $2.0\rearth$ planet with a 21.1~day period \citep{burke_et_al2014}. Our ARIES observations revealed a companion 5.42~\jmag and 5.28~\ksmag fainter than KOI~2914 at a distance of 3\farcs74. The predicted \kpmag of the companion is $Kp = 17.8$ ($\Delta Kp = 5.6$) if the star is a dwarf. KOI~2914 has also been observed with DSSI on WIYN, but the companion was outside the image area. The detected companion is likely to be the $J=16.64$ source found in UKIRT at a separation of 3\farcs95.  KOI~2914 does not exhibit a significant source offset during transit.

\subsection{KID~5473556 (formerly KOI~2939)}
KOI 2939 (KID 5473556) is an eclipsing binary with a single observed planetary transit \citep{welsh_et_al2012} and is no longer listed in the KOI catalog. We detected a companion 1.84~\ksmag fainter than KOI~2939 at a distance of 2\farcs78. The estimated \kpmag of the companion is $Kp = 15.8$ ($\Delta Kp = 2.2$). 

\subsection{KOI~2961}
 KOI~2961 hosts a single planet candidate with a radius of $1.2\rearth$ and an orbital period of 3.78~days \citep{burke_et_al2014}. Our ARIES observations revealed a companion 6.94~\ksmag fainter than KOI~2961 at a distance of 1\farcs95. The predicted \kpmag of the companion is $Kp = 21.5$ ($\Delta Kp = 8.9$) and the estimated dilution correction is only $0.01\%$ due to the large brightness contrast between KOI~2961 and the companion. 
 
 KOI~2961 has also been observed using speckle imaging with DSSI on WIYN at 692nm and 880nm. The companion we report in this paper was not detected in the 3\farcs2x3\farcs2 speckle image. At the distance of the companion, the 3-sigma detection limits for the speckle image were 4.04 magnitudes at 692nm and 3.953 magnitudes at 880nm. The lack of a detection in the speckle image is therefore unsurprising given the predicted faintness of the companion at bluer wavelengths. KOI~2961 does not display a significant source offset during transit. 

\subsection{KOI~2971}
This system contains a $0.8\rearth$ planet candidate with a 6.1~day period and a second $1.1\rearth$ planet candidate with a 31.9~day period \citep{burke_et_al2014}.  We detected a companion 6.69~\ksmag fainter than KOI~2971 at a distance of 3\farcs48. The predicted \kpmag of the companion is $Kp = 21.4$ ($\Delta Kp = 8.7$). KOI~2971 has also been observed using speckle imaging with DSSI on WIYN. The detected companion was identified in UKIRT and has a reported $J$ magnitude of 20.76. The detection limit near 3\farcs5 in our $J$~band ARIES image is $J=15.3$, so we are not able to estimate the $J$ band magnitude of the companion from our data. KOI~2971 does not exhibit a significant source offset during the transits of either KOI~2971.01 or 2971.02.
 
\subsection{KOI~2984}
KOI~2984 hosts a $1.1\rearth$ planet candidate with a 11.5~day orbit \citep{burke_et_al2014}. Our ARIES observations revealed a companion 3.81~\ksmag fainter than KOI~2984 at a separation of 3\farcs26. The estimated \kpmag of the companion is $Kp = 17.8$ ($\Delta Kp = 4.8$). KOI~2984 has also been observed using DSSI on WIYN and NIRC2 on Keck. The companion we detected was identified in UKIRT with a $J$ band magnitude of $16.0$. No closer companions were detected in the WIYN and NIRC2 images. KOI~2984 does not display a significant source offset during transit.

\subsection{KOI~3111}
This system hosts a $2.1\rearth$ planet candidate with a 10.8~day period and a $1.5\rearth$ planet candidate with a 4.3~day period \citep{burke_et_al2014}. We detected a companion 5.25~\ksmag fainter than KOI~3111 at a distance of 3\farcs33. The predicted \kpmag of the companion is $Kp = 19.4$ ($\Delta Kp = 6.5$). KOI~3111 has also been observed with speckle imaging using DSSI at WIYN. The companion we identified was visible in UKIRT and has a $J$ band magnitude of $17.98$. KOI~3111 exhibits a 2\farcs847 ($3.09\sigma$) source offset during the transits of KOI~3111.01, but only a 1\farcs89 ($1.64\sigma$) source offset during the transits of KOI~3111.02.

\subsection{KOI~3117}
KOI~3117 hosts a $1.5\rearth$ planet candidate with a 6.1~day period \citep{burke_et_al2014}. We detected a companion 6.1~\ksmag fainter than KOI~3117 at a separation of 2\farcs61. The estimated \kpmag of the companion is $Kp = 20.7$ ($\Delta Kp = 7.5$). KOI~3117 has also been observed with speckle imaging using DSSI at WIYN, but the source was not detected in the 3\farcs2 by 3\farcs2 field of view. The reported $3\sigma$ speckle detection limits for an annulus extending from 1\farcs7 - 1\farcs 9 (the farthest reported separation) are 4.095~magnitudes at 692nm and 3.381~magnitudes at 880nm. The lack of a speckle detection is not surprising given the large $Ks$ magnitude contrast between the target and the companion and the likelihood that the companion would be even fainter in the bluer 692nm and 880nm filters used in the speckle imaging. KOI~3117 does not display a significant source offset during transit.

\section{Detection Limits}
\label{sec:lim}
\begin{figure}[htbp]
\begin{center}
\centering
\includegraphics[width=0.5\textwidth]{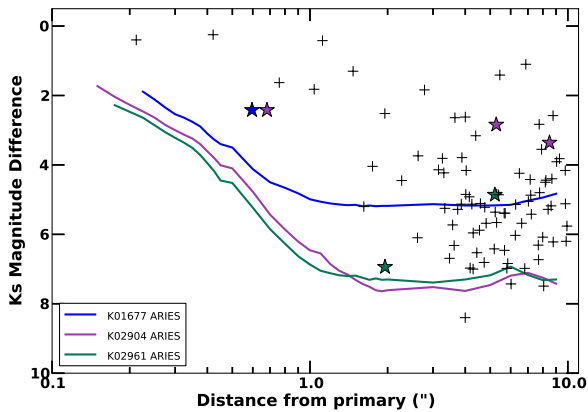}
\end{center}
\caption{Detected nearby stars (black crosses or stars) and detection limits (lines) on the presence of additional stars. We highlight the detected companions and detection limits for three systems: K01677 (blue), K02904 (magenta) and K02961 (green). K01677 is fainter than K02904 and K02961 by approximately 1.5 $Ks$ magnitudes. The stars in the vicinity of K00266, K00364, and K03242 were smeared by field rotation and are excluded from this plot because their magnitudes were underestimated. All of the stars detected around K00266, K00364, and K03242 are at separations of at least 3\farcs5 and were identified in UKIRT.}
\label{fig:detlims}
\end{figure}

In addition to measuring the brightness of companions, we calculated detection limits by measuring the total amount of flux in annuli centered on the target star. The widths of the non-overlapping annuli were $0\farcs05$ for separations within $0\farcs2$, $0\farcs1$ between $0\farcs2$ and $1\farcs0$, and $1''$ at separations beyond $1''$. We estimated the contribution from background stars by measuring the mean flux in an annulus with a radius of $10''$ and subtracted that background value from the total within each annulus to measure the flux due to the star at that distance. We then measured the standard deviation within each annulus and calculated the detection limit for each annulus as 5 standard deviations above the mean flux. For most targets we found a full width at half-maximum (FWHM) of 0\farcs25 and a limiting magnitude of $\Delta Ks = 5.3$ at $1''$.  However, under good conditions we are sensitive to companions as faint as $\Delta Ks = 7.5$ and as close as 0\farcs1 (see Figure~\ref{fig:detlims}). We provide detection limits for each target in Table~\ref{tab:detlim} and plot detection limits as a function of angular separation for three stars in Figure~\ref{fig:detlims}.

\LongTables
 \begin{deluxetable*}{clrccccccc}
\tablecolumns{10}
\scriptsize
\tablecaption{Limits on the Presence of Nearby Stars for All Observed Stars}
\tablehead{
\colhead{Object} &
\colhead{FWHM} &
\colhead{2MASS} &
\colhead{Companion} &
\multicolumn{6}{c}{Limiting $\Delta Ks$ for Annulus Centered At} \\
\noalign{\smallskip}
\cline{5-10}
\noalign{\smallskip}
\colhead{} &
\colhead{($''$)} &
\colhead{\emph{Ks}} &
\colhead{within $10''$} &
\colhead{0\farcs1} &
\colhead{0\farcs2} &
\colhead{0\farcs5} &
\colhead{1\farcs0} &
\colhead{2\farcs0} &
\colhead{4\farcs0} 
}
K00159	 &0.15	 &11.970	 &No 	& --  	& 2.3 	& 4.26 	& 6.15 	& 6.38 	& 6.33 \\
K00266	 &0.23	 &10.379	 &Yes 	& --  	& --  	& 3.0 	& 5.54 	& 6.36 	& 6.68 \\
K00330	 &0.32	 &12.384	 &No 	& --  	& --  	& 3.07 	& 4.82 	& 4.91 	& 4.92 \\
K00351	 &0.28	 &12.482	 &No 	& --  	& --  	& 3.24 	& 4.85 	& 4.82 	& 4.86 \\
K00364	 &0.1	 &8.645	 &Yes 	& 2.45 	& 3.68 	& 5.73 	& 8.03 	& 8.49 	& 8.6 \\
K00392	 &0.26	 &12.416	 &No 	& --  	& --  	& 3.26 	& 4.85 	& 5.01 	& 5.03 \\
K00664	 &0.25	 &12.001	 &No 	& --  	& --  	& 2.98 	& 4.98 	& 5.23 	& 5.38 \\
K00720	 &0.25	 &11.900	 &Yes 	& --  	& --  	& 3.35 	& 5.12 	& 5.25 	& 5.22 \\
K00886	 &0.49	 &12.648	 &No 	& --  	& --  	& 1.51 	& 3.0 	& 3.26 	& 2.95 \\
K00947	 &0.25	 &12.097	 &No 	& --  	& --  	& 3.48 	& 5.26 	& 5.42 	& 5.41 \\
K01219	 &0.26	 &12.469	 &No 	& --  	& --  	& 3.44 	& 5.34 	& 5.49 	& 5.51 \\
K01279	 &0.19	 &12.246	 &Yes 	& --  	& 1.84 	& 3.76 	& 5.16 	& 5.31 	& 5.33 \\
K01344	 &0.36	 &12.001	 &Yes 	& --  	& --  	& 2.27 	& 4.37 	& 4.73 	& 4.75 \\
K01677	 &0.29\tablenotemark{P}	 &12.687	 &Yes 	& --  	& --  	& 3.5 	& 4.99 	& 5.18 	& 5.17 \\
K01913	 &0.29	 &11.664	 &No 	& --  	& --  	& 3.01 	& 4.98 	& 5.19 	& 5.2 \\
K01977	 &0.23	 &11.551	 &Yes 	& --  	& --  	& 3.79 	& 5.79 	& 6.1 	& 6.13 \\
K02002	 &0.32	 &11.730	 &No 	& --  	& --  	& 2.91 	& 5.12 	& 5.35 	& 5.43 \\
K02072	 &0.28	 &11.970	 &No 	& --  	& --  	& 3.27 	& 5.28 	& 5.51 	& 5.51 \\
K02158	 &0.31	 &11.279	 &Yes 	& --  	& --  	& 3.24 	& 5.56 	& 5.78 	& 5.91 \\
K02159	 &0.3	 &11.982	 &Yes 	& --  	& --  	& 2.71 	& 4.73 	& 5.02 	& 4.92 \\
K02298	 &0.19	 &11.735	 &Yes 	& --  	& 1.86 	& 4.04 	& 5.57 	& 5.8 	& 5.95 \\
K02331	 &0.17	 &12.065	 &Yes 	& --  	& 2.12 	& 4.36 	& 6.05 	& 6.21 	& 6.18 \\
K02372	 &0.3	 &11.988	 &No 	& --  	& --  	& 2.4 	& 4.12 	& 4.54 	& 4.51 \\
K02399	 &0.22	 &12.187	 &Yes 	& --  	& --  	& 3.74 	& 5.27 	& 5.35 	& 5.32 \\
K02421	 &0.24	 &12.264	 &Yes 	& --  	& --  	& 3.59 	& 5.31 	& 5.55 	& 5.55 \\
K02426	 &0.3	 &12.199	 &Yes 	& --  	& --  	& 2.9 	& 4.53 	& 4.67 	& 4.79 \\
K02516	 &0.19	 &11.582	 &Yes 	& --  	& 1.74 	& 3.58 	& 5.55 	& 5.86 	& 5.85 \\
K02527	 &0.29	 &11.562	 &Yes 	& --  	& --  	& 2.96 	& 5.03 	& 5.4 	& 5.49 \\
K02581	 &0.22	 &11.808	 &No 	& --  	& --  	& 3.26 	& 5.23 	& 5.59 	& 5.57 \\
K02585	 &0.26	 &12.121	 &No 	& --  	& --  	& 2.97 	& 4.84 	& 5.03 	& 5.07 \\
K02623	 &0.18	 &12.075	 &Yes 	& --  	& 1.81 	& 3.48 	& 5.39 	& 5.76 	& 5.77 \\
K02672	 &0.35	 &10.285	 &Yes 	& --  	& --  	& 2.22 	& 4.75 	& 6.57 	& 6.91 \\
K02675	 &0.58	 &10.907	 &No 	& --  	& --  	& --  	& 3.5 	& 4.82 	& 5.5 \\
K02678	 &0.16	 &10.088	 &Yes 	& --  	& 2.52 	& 4.24 	& 6.77 	& 7.38 	& 7.64 \\
K02684	 &0.2	 &9.564	 &No 	& --  	& 1.49 	& 3.0 	& 5.25 	& 6.59 	& 7.09 \\
K02687	 &0.14	 &8.693	 &No 	& --  	& 2.11 	& 3.78 	& 5.97 	& 6.85 	& 6.8 \\
K02693	 &0.33	 &10.794	 &Yes 	& --  	& --  	& 2.61 	& 5.32 	& 6.21 	& 6.7 \\
K02706	 &0.16	 &9.109	 &Yes 	& --  	& 1.9 	& 3.46 	& 5.91 	& 7.67 	& 8.06 \\
K02720	 &0.19	 &8.996	 &No 	& --  	& 1.52 	& 3.07 	& 5.34 	& 7.01 	& 7.33 \\
K02722	 &0.23	 &11.993	 &Yes 	& --  	& --  	& 3.51 	& 5.43 	& 5.61 	& 5.62 \\
K02732	 &0.23	 &11.537	 &Yes 	& --  	& --  	& 3.59 	& 6.37 	& 6.81 	& 7.07 \\
K02754	 &0.08\tablenotemark{P}	 &10.627	 &Yes 	& --  	& 2.21 	& 3.95 	& 6.31 	& 7.05 	& 7.21 \\
K02755	 &0.13	 &10.706	 &No 	& --  	& 2.41 	& 4.26 	& 6.73 	& 7.42 	& 7.61 \\
K02771	 &0.38	 &10.462	 &Yes 	& --  	& --  	& 1.99 	& 4.51 	& 5.91 	& 6.59 \\
K02790	 &0.27\tablenotemark{P}	 &11.486	 &Yes 	& --  	& --  	& 3.28 	& 5.56 	& 6.05 	& 5.98 \\
K02792	 &0.36	 &9.761	 &No 	& --  	& --  	& 2.03 	& 4.38 	& 5.37 	& 5.81 \\
K02798	 &0.27	 &11.470	 &No 	& --  	& --  	& 3.33 	& 5.44 	& 5.68 	& 5.77 \\
K02801	 &0.18	 &9.472	 &No 	& --  	& 1.66 	& 3.25 	& 5.5 	& 6.04 	& 6.52 \\
K02803	 &0.2	 &10.642	 &Yes 	& --  	& 1.77 	& 3.43 	& 5.85 	& 6.53 	& 6.55 \\
K02805	 &0.39	 &11.919	 &No 	& --  	& --  	& 2.27 	& 4.85 	& 5.43 	& 5.71 \\
K02813	 &0.34	 &11.514	 &Yes 	& --  	& --  	& 2.51 	& 4.56 	& 5.36 	& 5.39 \\
K02829	 &0.19	 &11.444	 &Yes 	& --  	& 1.92 	& 4.01 	& 6.63 	& 7.14 	& 7.25 \\
K02833	 &0.15	 &11.123	 &Yes 	& --  	& 2.52 	& 4.64 	& 7.14 	& 7.52 	& 7.54 \\
K02838	 &0.44	 &11.857	 &Yes 	& --  	& --  	& 1.93 	& 4.28 	& 4.68 	& 4.85 \\
K02840	 &0.29	 &12.261	 &Yes 	& --  	& --  	& 3.32 	& 5.2 	& 5.4 	& 5.36 \\
K02859	 &0.25	 &12.052	 &No 	& --  	& --  	& 3.23 	& 5.32 	& 5.72 	& 5.76 \\
K02867	 &0.49	 &10.475	 &No 	& --  	& --  	& 1.64 	& 3.99 	& 5.3 	& 6.09 \\
K02879	 &0.16\tablenotemark{P}	 &11.099	 &Yes 	& --  	& 2.11 	& 4.0 	& 5.83 	& 6.5 	& 6.8 \\
K02904	 &0.18\tablenotemark{P}		 &11.359	 &Yes 	& --  	& 2.27 	& 4.1 	& 6.46 	& 7.61 	& 7.63 \\
K02913	 &0.23	 &11.361	 &Yes 	& --  	& --  	& 3.43 	& 5.83 	& 6.34 	& 6.4 \\
K02914	 &0.14	 &11.006	 &Yes 	& --  	& 2.84 	& 4.86 	& 7.32 	& 7.74 	& 7.92 \\
K02915	 &0.5	 &11.956	 &Yes 	& --  	& --  	& 1.67 	& 4.22 	& 4.72 	& 5.05 \\
K02916	 &0.28	 &12.402	 &No 	& --  	& --  	& 3.33 	& 4.73 	& 4.91 	& 4.89 \\
K02936	 &0.2	 &12.764	 &No 	& --  	& 1.99 	& 3.92 	& 4.91 	& 4.94 	& 4.93 \\
K02939	 &0.25	 &12.006	 &Yes 	& --  	& --  	& 3.4 	& 5.17 	& 5.28 	& 5.32 \\
K02948	 &0.38	 &10.322	 &No 	& --  	& --  	& 2.01 	& 4.4 	& 5.74 	& 6.42 \\
K02951	 &0.32	 &11.816	 &No 	& --  	& --  	& 2.96 	& 4.82 	& 4.95 	& 4.94 \\
K02961	 &0.16	 &11.290	 &Yes 	& --  	& 2.47 	& 4.52 	& 6.87 	& 7.3 	& 7.3 \\
K02968	 &0.42	 &10.293	 &No 	& --  	& --  	& 1.69 	& 3.91 	& 5.44 	& 5.97 \\
K02970	 &0.19	 &11.566	 &Yes 	& --  	& 2.01 	& 4.09 	& 6.76 	& 7.38 	& 7.29 \\
K02971	 &0.13	 &11.438	 &Yes 	& --  	& 2.67 	& 4.79 	& 6.87 	& 7.3 	& 7.34 \\
K02977	 &0.31	 &12.355	 &No 	& --  	& --  	& 2.75 	& 4.61 	& 4.75 	& 4.82 \\
K02984	 &0.3	 &11.637	 &Yes 	& --  	& --  	& 2.9 	& 5.6 	& 6.24 	& 6.42 \\
K03008	 &0.14	 &10.694	 &No 	& --  	& 2.58 	& 4.45 	& 6.93 	& 7.54 	& 7.76 \\
K03015	 &0.23	 &11.774	 &Yes 	& --  	& --  	& 3.48 	& 6.12 	& 6.58 	& 6.71 \\
K03017	 &0.22	 &11.757	 &No 	& --  	& --  	& 3.8 	& 5.75 	& 6.02 	& 6.02 \\
K03038	 &0.29	 &12.537	 &No 	& --  	& --  	& 0.14 	& 0.3 	& 0.63 	& 0.69 \\
K03060	 &0.26	 &11.554	 &No 	& --  	& --  	& 3.28 	& 6.12 	& 6.61 	& 6.7 \\
K03075	 &0.23	 &11.532	 &Yes 	& --  	& --  	& 3.7 	& 6.16 	& 6.75 	& 6.87 \\
K03083	 &0.28	 &11.401	 &No 	& --  	& --  	& 2.77 	& 5.31 	& 6.09 	& 6.43 \\
K03085	 &0.31	 &12.706	 &No 	& --  	& --  	& 2.93 	& 4.68 	& 4.85 	& 4.7 \\
K03097	 &0.5	 &10.649	 &No 	& --  	& --  	& 1.51 	& 3.79 	& 5.03 	& 5.78 \\
K03111	 &0.18	 &11.353	 &Yes 	& --  	& 2.16 	& 4.26 	& 6.62 	& 6.99 	& 7.05 \\
K03117	 &0.23	 &11.508	 &Yes 	& --  	& --  	& 3.65 	& 5.59 	& 5.87 	& 6.0 \\
K03122	 &0.14	 &10.819	 &Yes 	& --  	& 2.8 	& 4.8 	& 7.13 	& 7.88 	& 7.88 \\
K03128	 &0.51	 &11.900	 &Yes 	& --  	& --  	& --  	& 4.29 	& 4.88 	& 5.14 \\
K03242	 &0.1	 &11.520	 &Yes 	& 2.48 	& 3.59 	& 5.7 	& 7.96 	& 8.57 	& 8.6
 \enddata
  \tablenotetext{P}{FWHM determined using PSF fitting for stars with close companions.}
\label{tab:detlim}
\end{deluxetable*}

\begin{figure*}[tbhp]
\begin{center}
\centering
\includegraphics[width=1\textwidth]{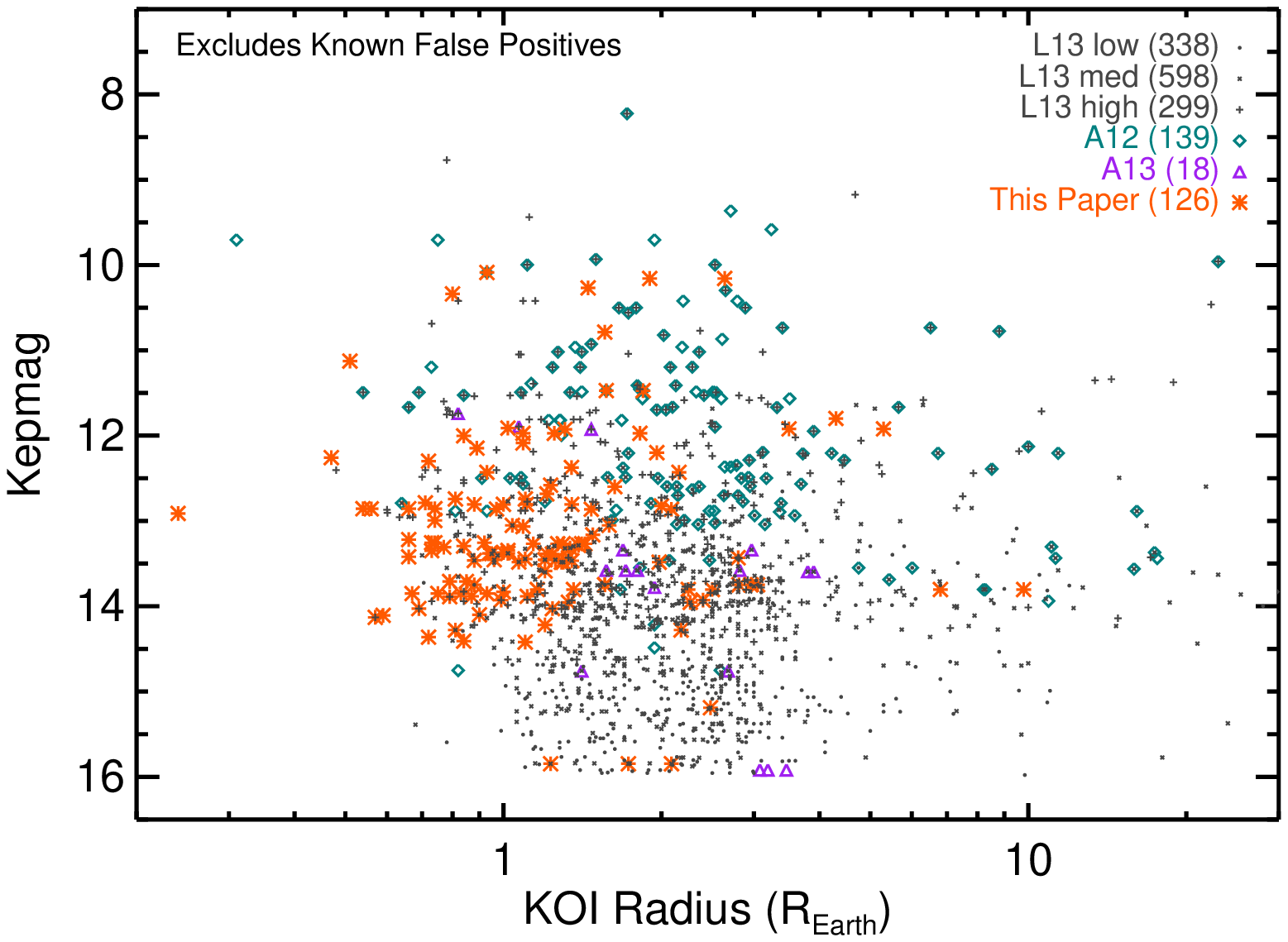}
\end{center}
\caption{\kepler magnitudes of host stars versus the radii of associated planet candidates for the KOIs observed by \citet{law_et_al2013} with Robo-AO (gray), \citet{adams_et_al2012} with ARIES and PHARO (teal diamonds), \citet{adams_et_al2013b} with ARIES (purple triangles), and in this paper with ARIES (orange stars). The symbols for the \citet{law_et_al2013} targets indicate the photometric quality of the observations as described in their Table~5. The KOI~radii were obtained from the cumulative planet candidate list at the NASA Exoplanet Archive\footnote{\url{http://exoplanetarchive.ipac.caltech.edu/cgi-bin/ExoTables/nph-exotbls?dataset=cumulative}} and have not been corrected for possible dilution due to the presence of nearby stars.}
\label{fig:surveys}
\end{figure*}

\section{Comparison to Previous Surveys}
\label{sec:disc}

\begin{figure}[tbhp]
\begin{center}
\centering
\includegraphics[width=0.5\textwidth]{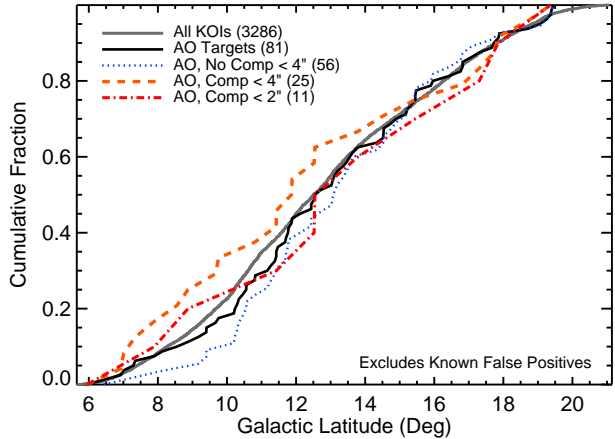}
\end{center}
\caption{Galactic latitude distribution of AO targets (black solid line), AO targets without identified companions (blue dotted line), AO targets with companions identified within $4''$ (orange dashed line), and AO targets with companions identified within $2''$ (red dot-dash line) compared to the galactic latitude distribution of all KOIs (gray). We have excluded all KOIs and AO targets that have been identified as false positives.}
\label{fig:gallat}
\end{figure}

\begin{figure*}[tbhp]
\begin{center}
\centering
\includegraphics[width=\textwidth]{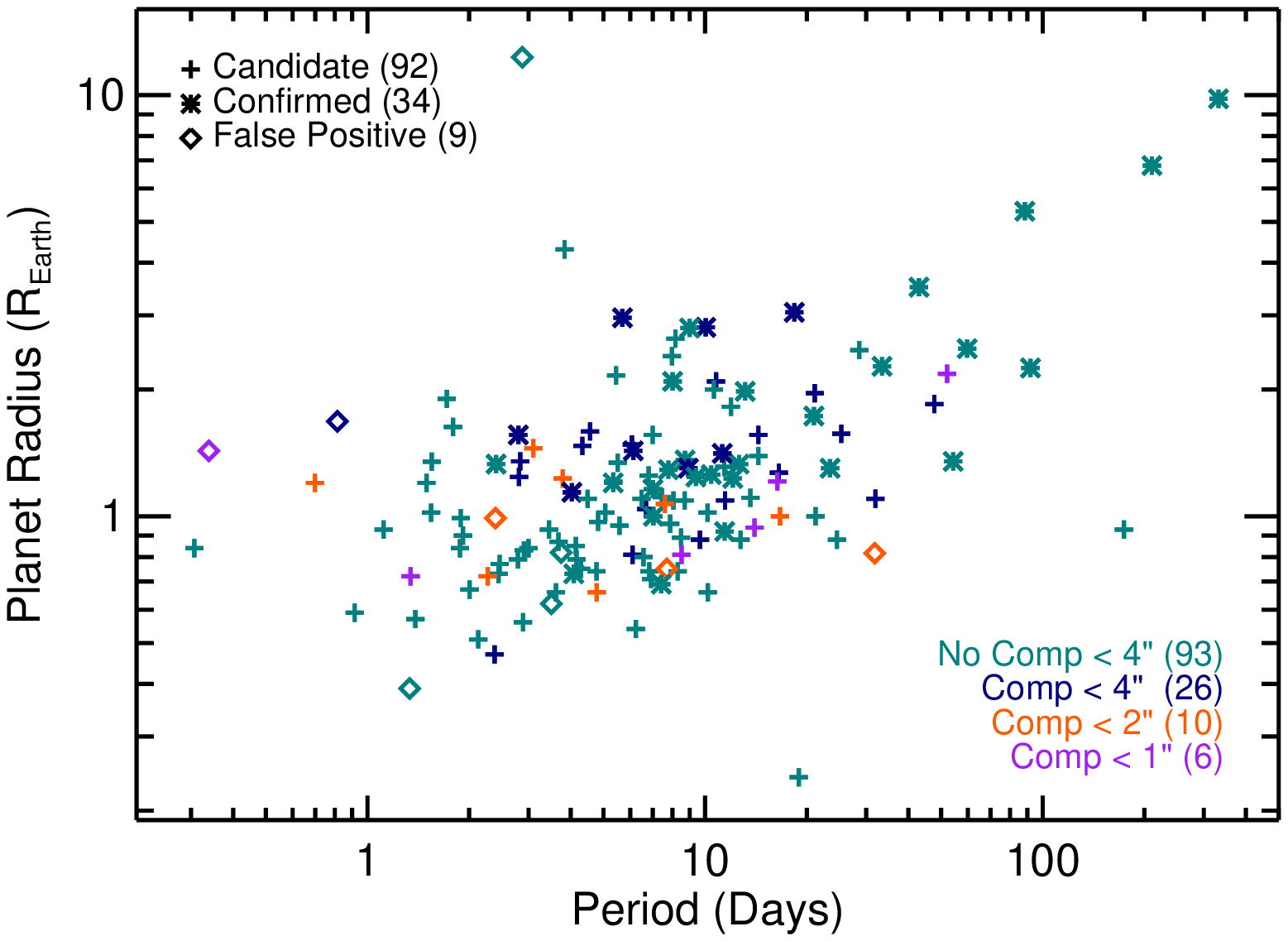}
\end{center}
\caption{Radii and periods for the planet candidates (crosses), confirmed planets (stars), and false positive KOIs (diamonds) orbiting the stars imaged in this study. Stars for which we did not detect a visual companion within $4''$ are shown in teal and stars with visual companions within $1''$, between $1-2''$, and between $2-4''$ are displayed in purple, orange, and navy, respectively. }
\label{fig:rpper}
\end{figure*}

As discussed in Section~\ref{sec:comp}, we detected visual companions within $2''$ around 11 of the \nhosts~targets that host planet candidates or confirmed planets. The overall companion rate of \percentwithintwo\% for planet (candidate) host stars is slightly lower than the rates of 20\% and 17\% found in \citet{adams_et_al2013b} and \citet{adams_et_al2012}, respectively and slightly higher than the rate of 7.4\% found by \citet{law_et_al2013} using Robo-AO on the robotic Palomar 60-inch telescope. Within $3''$ we found companions for 14 (17\%) of the \nhosts~targets hosting planet candidates or confirmed planets. This rate agrees well with the rate of 17\% found by \citet{lillo-box_et_al2012} using lucky imaging. 

Due to the efficiency of Robo-AO observations, the Robo-AO sample of 715~KOIs is much larger than the samples of 90, 12, 98, and 87~KOIs observed in \citet{adams_et_al2012}, \citet{adams_et_al2013b}, \citet{lillo-box_et_al2012}, and this paper, respectively. The Robo-AO team was therefore able to divide their sample into different categories and search for variations in the stellar multiplicity rate as a function of stellar or planetary properties. They found a slight ($1.6\sigma$) discrepancy between the stellar multiplicity of single KOI systems and multiple KOI systems, but the difference was not statistically significant. We also found a higher companion fraction for the 56 single KOI systems (18\%) compared to the 26 multiple KOI systems (4\%), which lends additional support to the theory that single KOI systems are more likely to be false positives than multiple KOI systems \citep{lissauer_et_al2012, lissauer_et_al2014}. %

Comparing the companion rates from different studies is not straightforward due to the small sample sizes of most of the studies and the differences in target sample selection, observing strategy, sensitivity, and weather conditions. The targets discussed in \citet{law_et_al2013} were selected randomly with the express goal of reproducing the general features of the full planet candidate population. In contrast, our observations and those of \citet{adams_et_al2012, adams_et_al2013b} were prioritized to target small planet candidates around bright or moderately faint ($Kp \lesssim14$) stars. As shown in Figure~\ref{fig:surveys}, the stars in the \citet{adams_et_al2012} sample are typically brighter than the stars observed by \citet{law_et_al2013} and the stars discussed in this paper. The median $Kp$ magnitude of the \citet{adams_et_al2012} sample is $Kp = 12.2$ whereas the median magnitude of our sample is $Kp = 13.3$. The \citet{law_et_al2013} sample extends to even fainter magnitudes and has a median magnitude of $Kp = 13.7$. 

In addition, the surveys reached different detection limits and operated in different bandpasses. In this paper and in \citet{adams_et_al2013b}, KOIs were observed in $J$ or $Ks$ using ARIES on the MMT. \citet{adams_et_al2012} presented $J$ and $Ks$ observations acquired with both ARIES on the MMT and PHARO on the Palomar Hale 200 inch telescope. \citet{lillo-box_et_al2012} conducted their observations in SDSS $i$ and $z$ bands using AstraLux on the 2.2~m telescope at Calar Alto Observatory. Finally, \citet{law_et_al2013} observed their targets at visible wavelengths using an SDSS-i' filter and a long-pass filter (LP600) that selects wavelengths redder than 600~nm and cuts off near 1000~nm. The shape of the LP600 filter matches the red end of the \kepler bandpass, so the contrast ratios measured in LP600 are more similar to the contrast ratios that the stars would have in the \kepler bandpass than the contrast ratios measured at near-infrared wavelengths.

In contrast, near-infrared observations are more sensitive to faint, red companions that may be below the detection limit at visible wavelengths. For example, the faint ($\Delta Ks = 2.5$, estimated $\Delta Kp = 3.1$) companion to KOI~2159 that we discuss in Section~\ref{ssec:2159} was classified as a ``likely'' Robo-AO detection rather than a ``secure'' detection because the detection significance was below their formal $5\sigma$ limit. Depending on weather conditions and the magnitude of the host star, observations with ARIES or PHARO may also have smaller inner working angles than Robo-AO observations. For instance, the close-in (0\farcs13) companion to KOI~1537 reported by \citet{adams_et_al2012} was too close to the target star to be resolved with Robo-AO. 

As shown in Figure~\ref{fig:gallat}, we find that the stars with visual companions are slightly more likely to be located at lower galactic latitudes than stars without identified companions. This indicates that some of the visual companions identified within $4''$ are likely to be background objects because the background density of stars is higher near the galactic plane. Concentrating on the stars with companions identified within $2''$, we see that the bias towards lower galactic latitudes is slightly reduced, as would be expected if many of the companions identified within $4''$ are not physically bound to the target star. Interestingly, none of our target stars have two detected companions within $4''$ whereas \citet{adams_et_al2012} detected multiple stars within $4''$ near eight of their 90~targets and \citet{lillo-box_et_al2012} found that 3\% of their targets had at least two companions within $3''$.

\section{Conclusions}
\label{sec:conc}
Our sample of target stars hosts \nsampcp~confirmed planets, \nsamppc~planet candidates, and \nsampfp~false positive KOIs. In Figure~\ref{fig:rpper} we display the radii and periods of these KOIs and denote which objects orbit stars with detected visual companions. Four of the stars with visual companions within $1''$ and all of the stars with companions within $2''$ host planet candidates smaller than $1.5\rearth$. In most cases, the estimated dilution corrections for these systems are small enough that the planet radii would change by only a few percent after accounting for the extra light in the aperture. In the extreme cases of KOI~2421 and KOI~2790, however, the approximate dilution corrections of 23\% and 17\%, respectively, would increase the radii of the associated planet candidates by over $0.15\rearth$. The change in the planet properties might be even larger if the planet candidates orbit the visual companions instead of the target stars. 

In addition to planet (candidate) host stars, our sample also includes \nfp~stars that were previously identified as planet host stars but have been revealed to be false positives. Our ARIES observations revealed visual companions within $4''$ of five of those stars (K02159, K02298, K02771, K02838, and K02879). Although this paper focuses on the search for companions to planet host stars, knowledge about the contamination of the light curves of stars without detected planets is equally important for computing planet occurrence rates. Most calculations of the frequency of planets (e.g., \citealt{cantanzarite+shao2011, youdin2011, howard_et_al2012, mann_et_al2012, traub2012, dressing+charbonneau2013, gaidos2013, kopparapu2013, petigura_et_al2013b, petigura_et_al2013, swift_et_al2013, morton+swift2014}; but see \citealt{fressin_et_al2013}) neglect flux contamination from nearby stars when estimating the smallest planet that could have been seen around a particular star, but additional light from a companion star could dilute the transit signals of small planets and render them undetectable. Failing to account for this dilution could therefore lead to an overestimate of the search completeness and an underestimate of the planet occurrence rate. In addition, stars with nearby visual companions of different spectral types might be misclassified due to their unusual colors, further complicating estimates of the search completeness.

For stars with companions closer than 2\farcs0, we estimated the appropriate dilution corrections for the radii of associated planet candidates. Depending on the magnitude differences and angular separations between the target stars and the identified companions, the approximate corrections to the planet radii varied from 0.2\% to 23\%. Given that radial velocity observations \citep{weiss_et_al2013, marcy_et_al2014, weiss+marcy2014} and planet formation models \citep{lopez+fortney2013}  have revealed that the transition between rocky and gaseous planets occurs at roughly $1.5\rearth$, this change has important implications for frequency of rocky planets in the galaxy. If the radii of many \kepler planet candidates are indeed underestimated, then the true frequency of rocky planets may be lower than previously estimated. However, we must caution that dilution from background stars will also make the detection of truly tiny planets more challenging. Accordingly, the \kepler census of rocky planets may be less complete than previously estimated. 

In the next few years, observations of \kepler target stars with Gaia \citep{perryman_et_al2001} will help disentangle blended systems by providing distance estimates for the host stars. In the case of blended systems in which the stars have nearly equal brightnesses, the distance reported by Gaia will be roughly 1.4~times that estimated from photometry alone. In that case, we will be able to infer that the system is a blend and that the radii of any planet candidates within the system are underestimated. Until we receive the Gaia data, however, we can inspect systems individually using ground-based observations like those presented in this paper and serendipitous space-based observations from the HST SNAP program (SNAP Program 12893; PI: R. Gilliland). 

\acknowledgments
{The authors gratefully acknowledge partial support from NASA grant NNX10AK54A. CD is supported by a National Science Foundation Graduate Research Fellowship. We thank David Ciardi for coordinating the \kepler Follow-up Observing Program and the anonymous referee for providing helpful suggestions to improve the paper. This research has made use of the \kepler Community Follow-Up Observing Program website (\url{https://cfop.ipac.caltech.edu}) and the NASA Exoplanet Archive, which is operated by the California Institute of Technology, under contract with the National Aeronautics and Space Administration under the Exoplanet Exploration Program. 

\emph{Facilities:} MMT (ARIES), \kepler}

\clearpage
\bibliography{../mdwarf_biblio.bib}

\end{document}